\documentclass[aip,pop,preprint]{revtex4-2} 

\usepackage[T1]{fontenc}
\usepackage[utf8]{inputenc}
\usepackage{graphicx}
\usepackage{amsmath,amssymb,bm}

\usepackage{hyperref}        

\def\be{\begin{equation}}
\def\bea{\begin{eqnarray}}
\def\eea{\end{eqnarray}}
\def\ee{\end{equation}}

\newcommand{\beq}{\begin{equation}}
\newcommand{\eeq}{\end{equation}}
\newcommand{\ba}{\begin{array}}
\newcommand{\ea}{\end{array}}
\providecommand{\apj}{ApJ}
\providecommand{\apjl}{ApJL}

\begin{document}

\title{Bridging the Gap between Collisional and Collisionless Plasma Shocks: A Simulation Study using OSIRIS}
\date{Submitted to \emph{Physics of Plasmas on 4 Aug 2025}}

\author{Yossef Nissim Kindi}
\email{ykindi@gmail.com}
\affiliation{Bar Ilan University, Ramat Gan, Israel, 5290002}

\author{Asaf Pe'er}
\affiliation{Bar Ilan University, Ramat Gan, Israel, 5290002}

\author{Antoine Bret}
\affiliation{ETSI Industriales, Universidad de Castilla-La Mancha,
             13071 Ciudad Real, Spain}
\affiliation{Instituto de Investigaciones Energ\'{e}ticas y Aplicaciones Industriales,
             Campus Universitario de Ciudad Real, 13071 Ciudad Real, Spain}

\author{Lu\'is O.~Silva}
\affiliation{GoLP/Instituto de Plasmas e Fus\~ao Nuclear,
             Instituto Superior T\'ecnico, Universidade de Lisboa,
             1049-001 Lisbon, Portugal}

\author{Kevin M. Schoeffler}
\affiliation{GoLP/Instituto de Plasmas e Fus\~ao Nuclear,
             Instituto Superior T\'ecnico, Universidade de Lisboa,
             1049-001 Lisbon, Portugal}
\affiliation{Institut f\"ur Theoretische Physik,
             Ruhr-Universit\"at Bochum, D-44780 Bochum, Germany}

\begin{abstract}
Shock waves in plasma can be characterized according to the physical mechanisms behind their formation. When collisions are frequent, shock dissipation is mainly binary-collision driven, and the shock width is often comparable to a few mean-free paths. In contrast, collisionless shocks rely on collective plasma processes to establish dissipation on scales far below the mean-free-path. The purpose of this study is to bridge these two regimes. Here, we present simulations using the OSIRIS particle-in-cell (PIC) code with the Coulomb collision module to explore the gradual transition between collisional and collisionless shocks by varying plasma parameters that determine collisionality. We show a smooth transition of the shock width in the intermediate region, where the ion plasma parameter is $N_D \approx 1$. Our numerical results confirm earlier theoretical predictions from existing theories in the asymptotic regimes, where the Mott–Smith ansatz combined with a full BGK collision operator was used to obtain the collisional‐regime shock width, and with the classical Tidman formalism that describes the collisionless limit. We demonstrate that the ion plasma parameter provides a useful metric for identifying when shocks transition from a fluid-like, mean-free-path scale to a collisionless, sub-mean-free-path scale. We discuss the importance of our results in astrophysical environments, where shock breakout causes a transition in the shock width, and marks the beginning of acceleration of particles to high energies.
\end{abstract}

\keywords{instabilities; plasmas; plasma nonlinear phenomena; shock waves; space plasma physics}

\maketitle

\section{Introduction}

Shock waves are commonly found in nature in systems that involve fluid flows. Such systems exist
on very different scales, ranging from the microphysical scale to the astronomical scale. As
a result, the properties of the shocks can vary considerably.
Conceptually, shock waves in plasma can be divided into two main categories: \emph{collisional} and \emph{collisionless}. In a collisional fluid, momentum and energy dissipation occur through frequent binary collisions of the constituent particles, causing shocks to form over scales of a few mean-free-paths. This phenomenon has been extensively studied in neutral fluids, and it remains valid as long as collisions are sufficiently frequent to thermalize particle velocities locally and drive the plasma to a near-Maxwellian distribution \citep{Zeldovich1967, Landau1987}.

In contrast, \emph{collisionless} shocks are sustained by collective interactions such as electrostatic or electromagnetic instabilities, wave-particle interactions, and self-regulated electric and magnetic fields \citep{Buneman1964,Sagdeev1966,Bale2003}. By definition, the collisional mean-free-path in a collisionless regime is very large compared to the system length scale; yet shocks can still form when upstream bulk kinetic energy is dissipated by collective plasma effects. Classic examples include the Earth's bow shock in the solar wind---measured to be orders of magnitude thinner than the proton mean-free-path \citep{Bale2003, Schwartz2011}, as well as shocks in supernova remnants and gamma-ray bursts \citep{Bamba2003, Bret2013}.

\citet{BretPeer2021} attempted to unify the understanding of collisional and collisionless shocks. They did that by using the Mott-Smith ansatz \citep{MottSmith1951} and the full BGK collision term \citep{BGK1954}  to derive a shock width in the collisional regime, and Tidman's approach \citep{Tidman1967} in the collisionless one. This approach describes a shock mediated by electrostatic turbulence on scales much smaller than the mean-free-path. 
In that work, \citet{BretPeer2021} identified the relevant regime using a single parameter that controls the transition, namely the ion plasma parameter, $N_D$:
\begin{equation}
N_D = \frac{4\pi}{3}\lambda_{Di}^{3}n_0 .
\end{equation}
Here, $n_0$ is the particle density and \(\lambda_{\mathrm{Di}}\) is the ion Debye length defined as  \(\sqrt{\frac{\varepsilon_0\,k_B\,T_i}{n_0\,q_e^2}}\), where $\varepsilon_0$ is the vacuum permittivity, $k_B$ the Boltzmann constant, $T_i$ the ion temperature and $q_e$ is the electron charge\footnote{We model a hydrogen (proton–electron) plasma, so the ion charge state is $Z=1$.}. Physically, \(N_D\) corresponds to the approximate number of particles inside one Debye sphere and it's directly related to the coupling strength of the plasma. For large $N_D > 1$, the collision rate becomes negligible compared to the plasma frequency, and kinetic energy dissipation must be mediated by collective plasma effects. On the other hand, for small $N_D$, binary collisions dominate, giving rise to fluid-like collisional behavior.

The approach used by \cite{BretPeer2021} was analytic, using the above mentioned assumptions, which are valid from the asymptotic limits of $N_D \gg 1$ and $N_D \ll 1$ up to the transition between the two regimes. It is therefore of interest to examine these assumptions, as well as provide proper calculations in the regime of $N_D \approx 1$.  
In this paper, we present a set of simulations that aim to \emph{bridge the gap} between these extremes. We employed the OSIRIS particle-in-cell (PIC) code \citep{Fonseca2002} to model two interpenetrating plasma streams with varying densities and temperatures, which in turn modify the effective collisionality. By tracking the shock formation and measuring the width of the density jump, we investigate the continuous transition from a collisional to a collisionless shock. We compare our simulation results with theoretical expectations, demonstrating how the plasma parameter $N_D$ offers a quantitative criterion to identify the shock's nature.

\section{Simulation Setup and the OSIRIS Code}

\subsection{OSIRIS model}
\label{sec:collisions-osiris}

Our numerical study uses the OSIRIS PIC code \citep{Fonseca2002, Fonseca2005}, a framework that evolves Maxwell's equations on a spatial grid and tracks the trajectories of discrete charged particles. These particles are subject to the Lorentz force and at each time step they are pushed to a new position and momentum according to the local electromagnetic (EM) fields inside the relevant cell. This movement of particles will, in turn, affect the physical quantities in the cell such as density, temperature, current, electric and magnetic fields. OSIRIS updates at each time step both the physical quantities and the particle's position and momentum in all the grid cells.

Since collisions are a crucial element in our study, we use the OSIRIS version that includes a collision module \citep{Fonseca2005, Fiore2010}. This module simulates collisions by randomly pairing particles within the same cell and assigning a scattering angle, at each time step. Each pair undergoes an elastic, energy and momentum conserving scattering in which the relative velocity $\mathbf{u}=\mathbf{v}_\alpha-\mathbf{v}_\beta$ is rotated by a small random angle $\theta$. The rotation angle $\theta$ is determined by defining a deflection parameter
\[
\delta \;=\;\tan\!\bigl(\tfrac{\theta}{2}\bigr).
\]
When $\delta \rightarrow 0$, the scattering angle approaches zero (no collisions), while $\delta \rightarrow \infty$ corresponds to a head-on encounter ($\theta \approx 180^\circ$). 

The deflection parameter $\delta$ is randomly chosen from a zero-mean Gaussian distribution, and its variance is proportional to the physical Coulomb collision frequency. For a pair of charges $q_\alpha$, $q_\beta$ and reduced mass $m_{\alpha\beta}=m_\alpha m_\beta/(m_\alpha+m_\beta)$ the variance per timestep $\Delta t$ is \cite[Eq.\,(8a)]{Takizuka1977}

\begin{equation}
\bigl\langle\delta^{2}\bigr\rangle \;=\;
\frac{q_\alpha^{\,2}q_\beta^{\,2}\,n_{L}\,\ln\Lambda}
     {8\pi\varepsilon_{0}^{2}\,m_{\alpha\beta}^{2}\,u^{3}}\,
\Delta t.
\label{eq:delta_var}
\end{equation}
Here, $n_{L}=\min(n_\alpha,n_\beta)$ is the lower of the two species densities, $\ln\Lambda$ is the Coulomb logarithm, and $u$ is the magnitude of $\mathbf{u}$.  
Equation~\eqref{eq:delta_var} ensures that the mean-square deflection accumulated in time $\Delta t$ is
$\langle\theta^{2}\rangle\propto\nu_{\alpha\beta}\Delta t$, when $\nu_{\alpha\beta}$ is the collision rate between the two species, reproducing the Landau collision operator in the small-angle limit \citep{Takizuka1977}.  Thus, although collisions are applied at every PIC step, their strength automatically scales with the physical collision frequency $\nu_{\alpha\beta}$.

From equation~\eqref{eq:delta_var} one sees that collisions become more frequent at a higher density and slower relative speeds between the particles, and the collision frequency scales inversely with the reduced mass. If more than two particles share a cell, they are grouped into pairs (or triplets, if an odd number remains) and scatter successively in one time-step.

Originally proposed by \citet{Takizuka1977}, this procedure has been refined for better computational efficiency by aggregating multiple small-angle collisions (within a single cell) into a single, larger-angle deflection \citep{Nanbu1997}, and, in addition has been extended to relativistic regimes \citep{Peano2009,Perez2012}. The OSIRIS code implements these adaptations \citep{Fiore2010}, thereby enabling self-consistent modeling of both electromagnetic fields and Coulomb collisions in large-scale PIC simulations.

\subsection{Simulation setup}
We perform simulations in a two-dimensional (2D) grid, with a varying number of cells, ranging from 3328 to 16640 in the x direction and 16 cells in the y direction (quasi one-dimensional simulation). The cells are resolved to the electron Debye length, i.e $\Delta x \approx \Delta y \approx 0.3 \lambda_{De}$, where $\lambda_{De}$ is the Debye length of the electrons. The initial electric and magnetic fields are zero in all simulations; while the initial temperature, density, and fluid velocities vary throughout the simulations, while all are kept non-relativistic. All our simulations consider electron-ion plasma, with a realistic mass ratio of 1836.
To ensure good accuracy, we use current smoothing and 36 particles per cell for both electrons and ions, namely $2 - 10 \times 10^6$ particles in total. The time steps of the simulation are measured in $1/\omega_p$, where $\omega_p$ is the plasma frequency defined as $\sqrt{\frac{n_0q_e^2}{m_e\epsilon_0}}$ where $m_e$ is the electron mass. All simulations ran for 4000 $1/\omega_p$, which we find to be a sufficient time for the formation of a shock.

To form shocks, we initialize two semi-infinite slabs \citep{Sorasio2006}. One slab is injected with a drift speed $v_0$ towards a stationary slab. To ensure the development of a shock, the drift speed is such as to ensure a Mach number larger than 1.  Table~\ref{tab:sim-parameters} lists the initial conditions of the simulation setups used. These include:
density of particles (number per ${cm}^3$, $n_0$); ratio between electron and ion temperature ($R_\mathrm{T}$); electron temperature in Kelvin ($T_e$); relative drift velocity between the two slabs ($v_0$) in units of the speed of light ($c$); the ion plasma parameter, $N_D$; and the Mach number of one slab in relation to another ($M$). The Mach number is calculated as $\frac{v_0}{v_s}$ when $v_s$ is the sound velocity of the incoming slab, defined by \citep{ChenBook}:
\begin{equation}
    v_s = \sqrt{\frac{k_B T_e + k_B \gamma T_i}{m_i}}.
\end{equation}
Here, $k_B$, $T_i$, $m_i$ and $\gamma$ are the Boltzmann constant, ion temperature, ion mass and the adiabatic index respectively. We further added to Table~\ref{tab:sim-parameters} the resulting shock width (SW) in units of the inertial length, $c/\omega_p$.

\begin{table}[h]
\centering
\footnotesize                
\setlength{\tabcolsep}{3pt}  
\caption{Parameters of the OSIRIS simulation runs performed. The parameters were chosen to represent different collisionalities . From left to right presented are the particle density $n_0$ with units of $cm^{-3}$; $R_T$, the temperature ratio between the electrons and ions ($T_e/T_i$); electron temperature $T_e$ in Kelvin; drift velocity of one slab with relation to another in units of speed of light $c$; plasma parameter $N_D$ of the ions; Mach number $M$; and resulting shock width (SW) in units of the inertial length $c/\omega_p$.}

\vspace{0.5em}
\begin{tabular}{c c c c c c c}
\hline
$n_0$ & $R_{\rm T}$ & $T_e$ & $v_0$ & $N_D$ & $M$ & SW \\
\hline
$1\times10^{22}$ & 400 & $2.97\times10^{7}$ & 0.0066 & $2.78\times10^{-1}$               & 3.99 & 0.50 \\   
$1\times10^{22}$ &   9 & $2.97\times10^{7}$ & 0.0066 & $8.25\times10^{1}$               & 3.67 & 0.73 \\   
$1\times10^{16}$ &   9 & $2.97\times10^{7}$ & 0.0066 & $8.25\times10^{4}$   & 3.67 & 0.77 \\   
$1\times10^{18}$ &   9 & $2.97\times10^{7}$ & 0.0066 & $8.25\times10^{3}$   & 3.67 & 0.88 \\   
$1\times10^{20}$ &   9 & $2.97\times10^{7}$ & 0.0066 & $8.25\times10^{2}$   & 3.67 & 0.66 \\   
$1\times10^{22}$ & 400 & $1.07\times10^{7}$ & 0.0040 & $6.01\times10^{-2}$  & 4.03 & 0.38 \\   
$1\times10^{20}$ & 400 & $1.07\times10^{7}$ & 0.0040 & $6.01\times10^{-1}$  & 4.03 & 0.29 \\   
$1\times10^{18}$ & 400 & $1.07\times10^{7}$ & 0.0040 & $6.01\times10^{0}$               & 4.03 & 0.29 \\   
$1\times10^{19}$ & 400 & $1.07\times10^{7}$ & 0.0040 & $1.90\times10^{0}$               & 4.03 & 0.24 \\   
$1\times10^{21}$ & 400 & $1.07\times10^{7}$ & 0.0040 & $1.90\times10^{-1}$  & 4.03 & 0.28 \\   
$1\times10^{20}$ & 400 & $2.97\times10^{7}$ & 0.0066 & $2.78\times10^{0}$               & 3.99 & 0.53 \\   
$1\times10^{18}$ & 400 & $2.97\times10^{5}$ & 0.0007 & $2.78\times10^{-2}$  & 4.23 & 0.06 \\   
$1\times10^{19}$ & 400 & $2.97\times10^{5}$ & 0.0007 & $8.81\times10^{-3}$  & 4.23 & 0.07 \\   
$1\times10^{16}$ & 400 & $2.97\times10^{7}$ & 0.0066 & $2.78\times10^{2}$   & 3.99 & 0.58 \\   
\hline
\end{tabular}
\label{tab:sim-parameters}
\end{table}

\textbf{Run procedure.} We allow each simulation to evolve until two well-defined shock fronts form near the interface between the two slabs. We then track the density profile along the transition from upstream to downstream plasma. The shock width $\ell$ is measured by locating where the density rises from $\sim10\%$ to $\sim90\%$ of the downstream density. This approach is repeated for the entire parameter space to produce a comprehensive view of how collisionality affects shock structure.

\section{Results: shock widths in collisional and collisionless plasmas}

\subsection{Shock wave formation}

\begin{figure}[ht]
\centering
\includegraphics[width=0.45\textwidth]{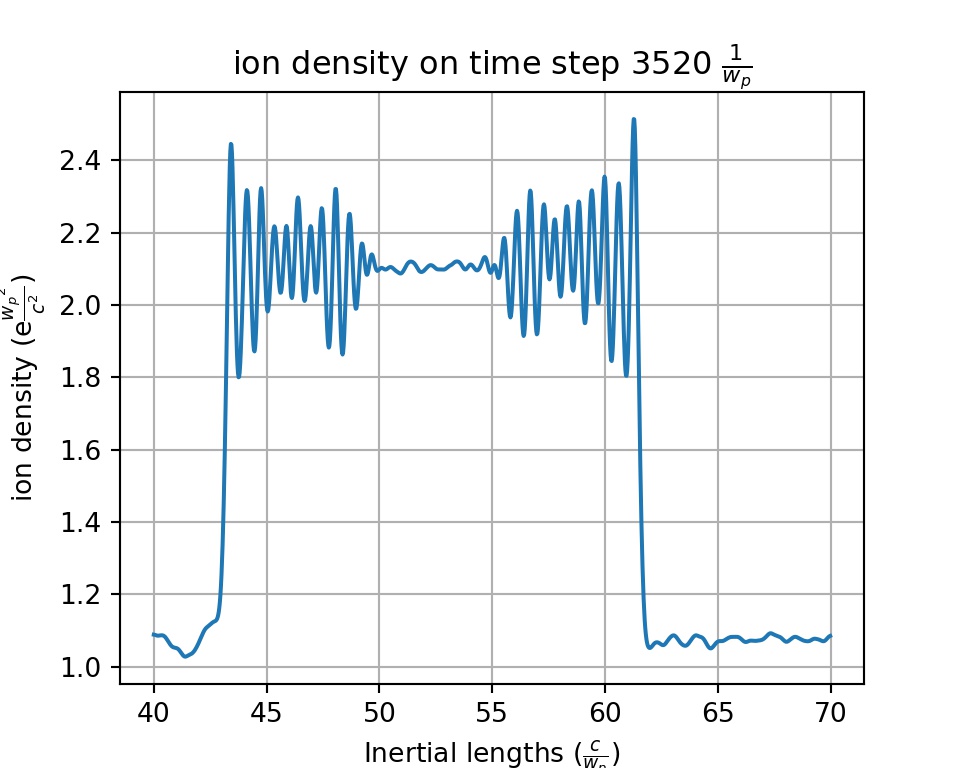}
\caption{Ion density profile from an OSIRIS simulation in the collisional regime.}
\label{fig:ion-collisional}
\end{figure}

We identify the formation of shock waves via the abrupt change in the ion density, resulting from the interpenetration of the plasma slabs at a supersonic speed.
Figure~\ref{fig:ion-collisional} illustrates
the ion density profiles taken from our OSIRIS simulation in the collisionality intermediary regime, with $N_D=2.78$. The two shocks around $43 c/\omega_p$ and $62 c/\omega_p$ are clearly visible. In addition, the density oscillations in the downstream of the shocks that are typical of electrostatic collisionless shocks \citep{Forslund1970} are clearly seen. These occur as a result of the conversion of kinetic energy to electromagnetic waves, increasing their amplitude.

\subsection{Plasma parameter as an indication of the shock collisionality}
We quantify plasma collisionality using the ion plasma parameter \(N_D\). A large \(N_D\) signifies that collective plasma interactions overshadow binary collisions, leading to a more \emph{collisionless} behavior, whereas small \(N_D\) indicates a stronger, \emph{collisional} coupling. We estimate the ion collision rate as \(\nu_{\mathrm{coll}} = v_{\mathrm{th}}/l_{\mathrm{mfp}}\), where \(v_{\mathrm{th}}\) is a characteristic thermal velocity defined as \( \sqrt{\frac{k_B\,T_i}{m_i}} \) and \(\ell_{\mathrm{mfp}}\) is the mean-free-path. Here we consider ion-ion collisions only, so $\nu_{coll}$ and $l_{mfp}$ refer to ion-ion scattering. Our mean-free-path extended to $N_D < 1$ is derived from \citet{Daligault2012} (their equation 2),
\begin{equation}
  \ell_{\mathrm{mfp,ion}}
  \;\approx\;
  24.655\frac{N_D \,\lambda_{\mathrm{D}}}{\ln(1 + 6.477 ~N_D\ )}.
\end{equation}

Hence, omitting the numerical constants for clarity, the collision rate normalized by the plasma frequency scales as
\[
\frac{\nu_{\mathrm{coll}}}{\omega_{p}}
\;\sim\;
\frac{\ln\!\bigl[1+N_D\bigr]}{N_D}.
\]
For \(N_D \ll 1\), one finds \(\ln(1+N_D)\approx N_D\) that implies a ratio \(\sim 1\), indicative of a strongly collisional regime. Conversely, at large \(N_D\), the factor \(\ln(1+N_D)/N_D\approx N_D^{-1}\), indicating that collisions are negligible compared to collective oscillations, thus identifying a collisionless regime. This direct dependence on \(N_D\) demonstrates why the plasma parameter is an excellent metric for gauging collisionality. 

\subsection{Shock width dependence on the plasma parameter}

To vary the ion plasma parameter we scan the density $n_0$ together with the ion temperature $T_i$. Individually, neither $n_0$ nor $T_{i}$ showed a systematic correlation with the shock width; only their combined influence through $N_D$ \footnote{We numerically checked that a four orders of magnitude change in the density between two points with similar plasma parameters result in a similar shock width. Similarly, a two orders of magnitude variation in the temperature does not affect the shock width for similar plasma parameter.}.
Following \citet{Sorasio2006} and \citet{Fiuza2013} we initialize hot electrons and cold ions, a choice that preserves an electron‐dominated sound speed $c_s\simeq\sqrt{k_B T_e/m_i}$ and therefore speeds up shock formation.  In the collisional regime ($N_D\lesssim1$) we keep $T_e/T_i\!\approx\!400$, which leaves $c_s$ governed by electrons while retaining a finite $T_i$ that produces measurable ion–ion scattering.  Achieving the collisionless limit demands $N_D\gg1$; since $N_D\propto T_i^{3/2}$, we raise $T_i$ (and hence lower the ratio) to $T_e/T_i\!\approx\!9$.  This still ensures electron control of $c_s$ yet boosts $T_i$ enough to reach $N_D\gtrsim10^{2}$, yielding a genuinely collisionless shock.

\begin{figure}[htbp]
\centering
\includegraphics[width=0.45\textwidth]{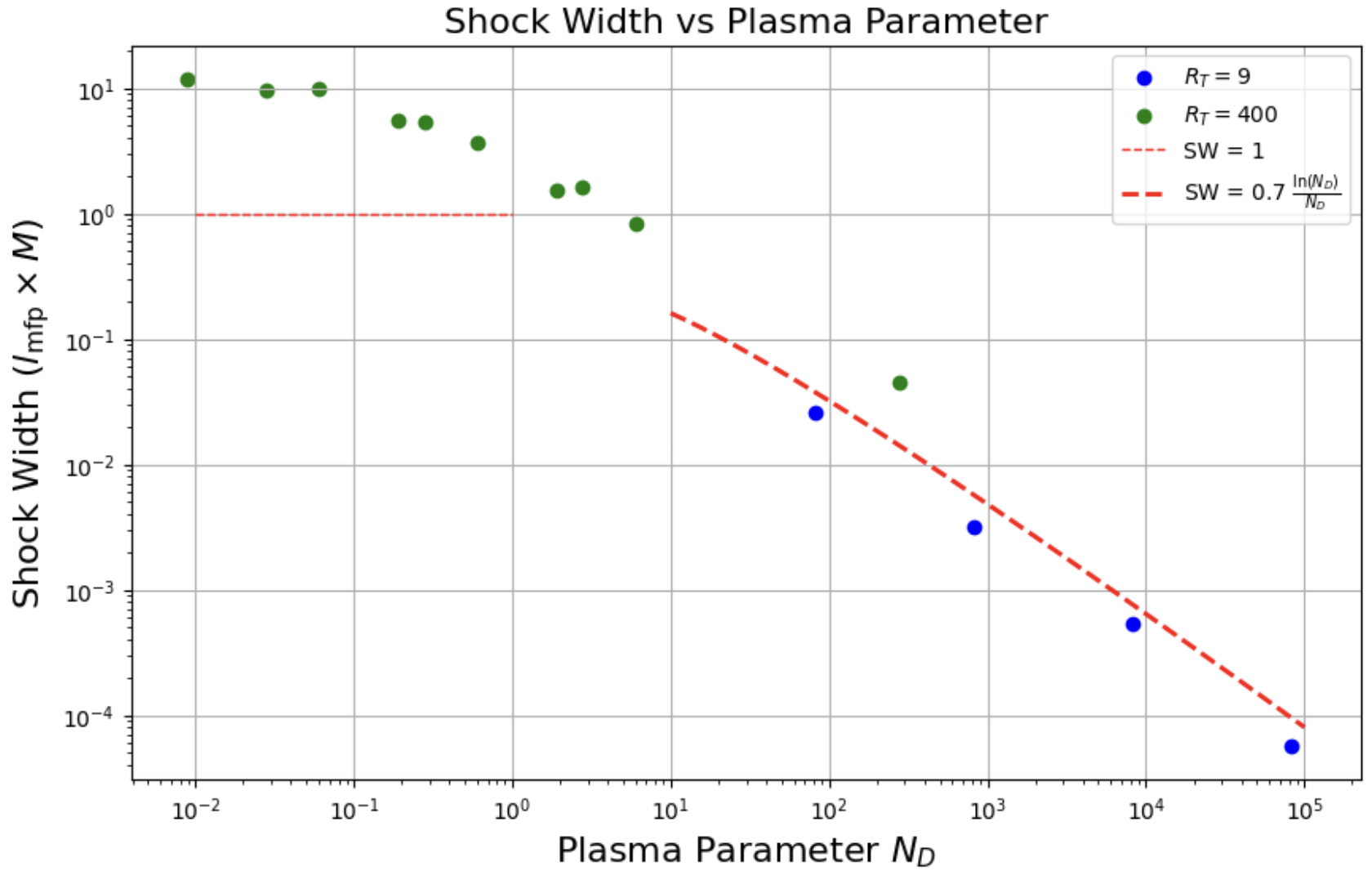}
\caption{Measured shock width (in units of the upstream mean-free-path times the Mach number) vs.\ the ion plasma parameter $N_D$. Data points are from multiple runs spanning densities $10^{16}\text{--}10^{22}$ cm$^{-3}$ and various temperature ratios. Around $N_D \sim 1$, shock widths begin decreasing below the collisional scale, indicating a shift to collisionless mediation. The horizontal dashed line $\ell/(M \cdot l_{\mathrm{mfp}})=1$ and the slanted dashed curve $\ell/(M \cdot l_{\mathrm{mfp}})=0.7\,\ln{N_D}/N_D$ are the collisional and collisionless scalings, respectively, predicted by \citet{BretPeer2021}.
}
\label{fig:shock-width-plot}
\end{figure}

In Figure~\ref{fig:shock-width-plot}, we plot the measured shock widths, normalized by the upstream mean-free-path $l_{\mathrm{mfp}}$ {and the Mach number}, as a function of the upstream ion plasma parameter $N_D$. Each data point corresponds to a distinct OSIRIS run, as is given in Table~\ref{tab:sim-parameters}. At low $N_D$ (strong collisional coupling), the shock widths is nearly independent on the value of $N_D$, and is of the order of a few $l_{\mathrm{mfp}}$. This is the expected result in the collisional regime \citep{BretPeer2021}.
At higher values of $N_D$, $N_D>1$, the shock width drops below $l_{\mathrm{mfp}}$, and is $\propto N_D^{-1}$, as expected theoretically \citep{Tidman1967}. This result is consistent with the expectation in the collisionless regime.
In the intermediate ($N_D \approx 1$) regime, we find a smooth, gradual transition between these two asymptotic regimes.

In Figure~\ref{fig:shock-width-plot} we also plot the asymptotic scalings derived by \citet{BretPeer2021}.  
The horizontal dashed line $\ell/(M \cdot l_{\mathrm{mfp}})=1$ and the slanted dashed curve
$\ell/(M \cdot l_{\mathrm{mfp}})=A\,\ln{N_D}/N_D$, 
where the dimensionless prefactor \(A\) was not fixed in the original theory. A least–squares fit to our  measurements yields \(A\simeq0.7\); the red dashed line in the figure adopts this value.  
The measured points track the two analytic branches in their respective limits and exhibit a smooth bridge around \(N_D\sim1\), thereby validating the crossover scenario proposed by \citet{BretPeer2021}.

\section{Analysis of Results and Comparison with Theory}
Our findings are aligned with previous theoretical works, and expand them to the intermediate region, where analytical approximations are invalid. In particular, \cite{BretPeer2021} proposed that the ion plasma parameter $N_D$ can act as a toggle for shock dissipation mechanisms: for $N_D<1$, collisions dominate and the Mott-Smith formalism \citep{MottSmith1951} with the BGK collision term is appropriate; for $N_D>1$, one transitions smoothly to a collisionless shock picture reminiscent of Tidman's analysis \citep{Tidman1967}.

The Mott-Smith approach treats the ion distribution function as a superposition of upstream and downstream drifting Maxwellians, with the shock thickness scaling as a few mean-free-paths \citep{MottSmith1951}. Our simulation data at low plasma parameter $N_D$ follow this pattern, with $\ell/l_{\mathrm{mfp}}$ in the range of a few mean-free-paths. This is consistent with the fact that the shock formation is driven mainly by binary collisions.

Tidman \citep{Tidman1967} analyzed high-Mach-number electrostatic shocks, deriving that the width scales with the ion Debye length (multiplied by factors dependent on the Mach number and logarithmic corrections). In the collisionless limit, $l_{\mathrm{mfp}}$ becomes irrelevant, so $\ell \ll l_{\mathrm{mfp}}$. Our high-$N_D$ runs exhibit precisely such behavior; we detect narrower shock fronts, often accompanied by electrostatic potential barriers that reflect a noticeable fraction of incoming ions. 

Thus, the bridging scenario proposed by Bret \& Pe'er (2021) is borne out in our simulations: the plasma parameter $N_D$ is indeed a reliable metric for indicating when to expect collisionless or collisional behavior, at least in this unmagnetized, nonrelativistic context.

\section{Conclusion}
We have presented a series of 2D PIC simulations with the OSIRIS code, in which two interpenetrating electron-ion plasmas form shocks across a broad spectrum of collisionalities. By systematically varying density, temperature, and relative flow velocity, we probed the transition from collisional to collisionless shock formation. Our results corroborate recent theoretical predictions that this transition is governed by the ion plasma parameter, $N_D$. When $N_D < 1$, the shock width is controlled by classical collisional processes, matching Mott-Smith-like theories. Above $N_D \approx 1$, collisionless collective effects dominate, yielding narrower shocks consistent with Tidman's work on electrostatic shocks. 

Several caveats accompany our study. First, although OSIRIS solves the full set of electromagnetic Maxwell equations, we operate in a quasi-1D, low-Mach-number regime where electrostatic modes are expected to dominate; under these conditions we can safely neglect the small electromagnetic corrections. The simulations also remain two-dimensional and unmagnetised, so external fields and fully three-dimensional electromagnetic instabilities—which may modify the transition—are deliberately excluded. Second, we adopt a fixed electron‐to‐ion temperature ratio ($T_e/T_i\simeq400$ for collisional runs and $\simeq9$ in the most collisionless cases) to keep the sound speed electron–dominated; alternate choices could shift the precise value of $N_D$ at which the crossover occurs. Future work should therefore relax these same simplifications: extend the calculations to fully three-dimensional, magnetized geometries so as to capture electromagnetic instabilities, scan a wider range of temperature ratios (and hence of the sound speed), and include additional physics such as multiple ion species or relativistic drifts—each of which will test whether the 
$N_D$ scaling inferred here remains robust once the neglected effects are reinstated. These extensions would also complement, and extend into the non-relativistic limit, the relativistic unmagnetized-shock studies in which the Weibel instability governs the transition (\citealp{Fiore2010}).

The collisional--collisionless crossover studied here naturally occurs when a dense shock leaves the stellar interior and enters a much lower--density medium \citep{Bret2018}.  
Numerical models of trans-relativistic supernova shock breakouts, for instance, show that the front is collisional while still inside the envelope but turns collisionless immediately after breakout \citep{Kashiyama2013}, once the upstream density drops and the ion plasma parameter $N_D$ surges.  
A similar evolution is inferred for coronal mass-ejection (CME) shocks in the solar corona: radio and white-light diagnostics indicate a collisional front at $R\!\lesssim\!6\,R_\odot$, which is replaced by a collisionless discontinuity beyond $R\!\gtrsim\!10\,R_\odot$ as the proton mean-free-path exceeds the shock width \citep{Eselevich2012}.  
In that newly collisionless phase the self-generated electromagnetic fields can accelerate a fraction of upstream ions to non-thermal energies, producing high-energy radiation and neutrinos \citep{Kashiyama2013}.  
By demonstrating that the shock width obeys a universal $N_D$ scaling across the transition, our simulations provide a practical diagnostic for pinpointing where such particle acceleration should be triggered in both astrophysical outflows and laboratory experiments that emulate breakout conditions.

\begin{acknowledgments}
We gratefully acknowledge helpful discussions with Lorenzo Sironi, whose insights
helped us finishing this project.
Antoine Bret also acknowledges support by the Ministerio de Economía y Competitividad of Spain
(Grant No. PID2021-125550OB-I00).
Kevin M. Schoeffler is supported by the German Science Foundation DFG within the Collaborative
Research Center SFB1491.
\end{acknowledgments}

\section*{Data Availability}
All simulation input files and analysis scripts are available from the corresponding author upon request.


\bibliographystyle{aipnum4-2}
\bibliography{references}

\end{document}